\documentclass{amsart}
\usepackage[english]{babel}
\usepackage{multicol}
\usepackage{epsfig}
\usepackage{amsfonts}
\usepackage{amssymb}
\usepackage{latexsym}
\usepackage{amsmath}
\usepackage{amsthm}
\usepackage{tikz}
\usepackage{pgf}
\usepackage{url}
\usepackage{a4wide}

\usetikzlibrary{calc,backgrounds, arrows, automata}

\numberwithin{equation}{section}

\newtheorem{theorem}{Theorem}[section]

\newtheorem*{lemma*}{Division Lemma}

\newtheorem{proposition}[theorem]{Proposition}
\newtheorem{procedure}[theorem]{Procedure}
\theoremstyle{definition}

\theoremstyle{remark}

\newtheorem{example}[theorem]{Example}

\def\mapping#1{{#1}^\prime}

\begin{document}

\title{Generating a chain of maps which preserve the same integral as a given map}

\author{J.M. Tuwankotta$^\dagger$, P.H. van der Kamp$^\ddagger$, G.R.W. Quispel$^\ddagger$,
and K.V.I. Saputra$^\diamond$}

\thanks{$^\dagger$ Analysis and Geometry Group, FMIPA, Institut Teknologi Bandung, Ganesha no. 10, Bandung, Indonesia.  E-mail: theo@math.itb.ac.id. }

\thanks{$^\ddagger$ School of Mathematical and Statistical Sciences, La Trobe University, Bundoora, Vic. 3083, Australia}

\thanks{$^\diamond$ Applied Math, Universitas Pelita Harapan, Jl. M.H. Thamrin Boulevard, Tangerang, 15811 Banten, Indonesia}

\begin{abstract}
We generalise the concept of duality to systems of ordinary difference equations (or maps). We propose a procedure to construct a chain of systems of equations which are dual, with respect to an integral $H$, to the given system, by exploiting the integral relation, defined by the upshifted version and the original version of $H$.  When the numerator of the integral relation is biquadratic or multi-linear, we point out conditions where a dual fails to exists.  The procedure is applied to several two-component systems obtained as periodic reductions of 2D lattice equations, including the nonlinear Schr\"{o}dinger system, the two-component potential Korteweg-De Vries equation, the scalar modified Korteweg-De Vries equation, and a modified Boussinesq system.
\end{abstract}

\maketitle


\markboth{  }{  }

\section{Introduction}
Discrete dynamical systems arise quite naturally in applications, for example as approximations for systems of ordinary or partial differential equations.  In some cases, the original system has special properties so that one would like to have a discretisation which preserves some (if not all) of those properties.   Integrability is one of the properties that one would like to preserve in the discretisation.  For examples of discretization of integrable partial differential equations such as the (modified) Korteweg-De Vries (MKdV) or the sine-Gordon equations, while preserving integrability, see \cite{Ablowitz, Hirota, QuispelNijhoffCapelvanderLinden}.

Discrete integrable systems are interesting due to their richness in structure. In the literature,  integrable ordinary difference equations,  both autonomous
  \cite{Papageorgiou, QuispelCapelPapageorgiouNijhoff} (e.g. the QRT maps \cite{QuispelRobertsThompson1,QuispelRobertsThompson2}) and non-autonomous (e.g. discrete Painleve equations \cite{Gramaticos}) have been extensively studied.   In 2010, J.J. Duistermaat published a wonderful book with the title Discrete Integrable Systems: QRT Maps and Elliptic Surfaces \cite{Duistermaat}, where algebraic geometry and complex analytic geometry have been used to derive the properties of the QRT map.

In \cite{Quispel}, the concept of duality for discrete $d$-th order ordinary difference equations (O$\Delta$E) was introduced. Let a $d$-th order ordinary difference equation with a number of integrals be given.   The idea of constructing a dual system in \cite{Quispel} is based on factoring out the so-called integrating factor from the difference between the upshifted and the original linear combination of the integrals of the given equation.   Thus, the dual equation can be seen as a new difference equation which preserves the same linear combination of integrals. In the recent paper \cite{DAKP}, the concept of duality is extended to lattice equations, by exploiting characteristics of conservation laws, and a 3D lattice equation dual to the lattice AKP equation is derived.

Generally speaking, dual equations to integrable equations do not need to be integrable themselves. However, the existence of one or more integrals (for O$\Delta$Es), or conservation laws (for P$\Delta$Es), is guaranteed. As these properties are somewhat special, dual equations are good candidates for new integrable equations. In \cite{Quispel}
dual equations to $(d-1,-1)$-periodic reductions\footnote{$(s_1,s_2)$-periodic reductions were first introduced in \cite{Papageorgiou,QuispelCapelPapageorgiouNijhoff}, cf. \cite{IVPs,vanderKamp}. By introduction a new independent variable $k=s_2m-s_1n$ the periodicity condition $k(m,n)=k(m+s_1,n+s_2)$ is satisfied; it is a discrete travelling wave reduction where a lattice equation gives rise to an O$\Delta$E.} of the modified Korteweg-De Vries (mKdV) lattice equation are shown to be integrable maps, namely level-set-dependent mKdV maps. In \cite{DTKQ}, $\lfloor \frac{d-1}{2}\rfloor$ integrals are provided explicitly for a novel hierarchy of maps dual to the linear equation $u_{n}=u_{n+d}$. The integrability of these maps is established in \cite{HKQ}. In \cite{DAKP} it is shown that reductions of the dual AKP equation include Rutishauser's quotient-difference (QD) algorithm, the higher analogue of the discrete time Toda (HADT) equation and its corresponding quotient-quotient-difference (QQD) system \cite{HA}, the discrete hungry Lotka-Volterra system, discrete hungry QD, as well as the hungry forms of HADT and QQD \cite{hHA}. A relation with the BKP equation and hence its integrability was pointed out by Schief, cf. \cite{DAKP2}.

So far, in the literature, duality has only been applied to scalar equations. For a scalar equation (O$\Delta$E) $E=0$ with an integral
$H$ we have that $\Delta H=H_{n+1}-H_n=E\Lambda$ factorises, and naturally the dual of the equation $E=0$ is given by $\Lambda=0$. In this paper we generalise the concept of duality to systems of equations. The intrinsic problem for a system of equations, such as $E_1=E_2=0$, is that the difference $\Delta H = f(E_1,E_2)$ does not factorise. For example, the following system of two difference equations, where sub-indices on $x,y$ denote the number of shifts,
\begin{equation} \label{intexa}
\begin{split}
E_1&=\left( x_{{2}}y_{{0}}-x_{{3}}y_{{1}} \right) p+ \left(x_{{3}}y_{{2}} -x_{{1}}y_{{0}} \right) q =0,\\
E_2&=\left( x_{{0}}y_{{1}}y_{{3}}-x_{{1}}y_{{0}}y_{{2}} \right) p+ \left(
x_{{2}}y_{{0}}y_{{1}} -x_{{0}}y_{{2}}y_{{3}}\right) q=0
\end{split}
\end{equation}
admits an integral given by
\[
H=\left( {\frac {y_{{1}}}{y_{{0}}}}+{\frac {x_{{1}}}{x_{{2}}}}+{\frac {
x_{{0}}}{x_{{1}}}}+{\frac {x_{{2}}y_{{1}}}{y_{{2}}x_{{1}}}}+{\frac {y_
{{2}}}{y_{{1}}}}+{\frac {x_{{1}}y_{{0}}}{x_{{0}}y_{{1}}}} \right) p-
 \left( {\frac {x_{{0}}}{x_{{2}}}}+{\frac {x_{{2}}y_{{0}}}{x_{{0}}y_{
{2}}}}+{\frac {y_{{2}}}{y_{{0}}}} \right) q,
\]
and we have (the coefficients $c_i$ are given in Appendix A)
\begin{equation} \label{cs}
\Delta H = \frac{
c_1 E_{{1}}
+c_2 E_{{2}}
+c_3 E_{{1}}E_{{2}}
+c_4 {E_{{1}}}^{2}
+c_5 {E_{{2}}}^{2}
+c_6 E_{{1}}{E_{{2}}}^{2}
}{y_{{0}}y_{{1}}y_{{2}}x_{{2}}x_{{1}} \left( px_{{1}}y_{{0}}y_{{2}}-qx_{
{2}}y_{{0}}y_{{1}}+E_{{2}} \right)  \left( px_{{2}}y_{{0}}-qx_{{1}}y_{
{0}}-E_{{1}} \right) x_{{0}}},
\end{equation}
which does not factorise.

We propose the following solution to this problem. For a system with $N$ equations,
one constructs $N$ duals by setting $E_j=0$ for all $1\leq j\neq i \leq N$
for each $i$. In each case, the remaining expression $E_i$ will factor out, $\Delta H = E_i \Lambda_i$, and a dual system is obtained which has the form
\[
E_1=E_2=\cdots=E_{i-1}=\Lambda_i=E_{i+1}=\cdots=E_N=0.
\]
For the above example \ref{intexa}, the two duals this method yields are, for $i=1$,
\[
E_1=0,\quad c_2 +c_5 E_{{2}}=0
\]
and, for $i=2$,
\[
E_2=0,\quad c_1+c_4 E_{{1}}=0,
\]
where $c_1,c_2,c_4,c_5$ are given in Appendix A.
Clearly both these systems preserve the integral $H$.

One can continue taking duals of duals, and by doing so one may obtain either finitely or infinitely many dual systems. Starting from an integrable system with several integrals the following questions arise
\begin{itemize}
\item[Q1] for which (linear) combinations of integrals are the dual systems integrable?
\item[Q2] are the dual systems to the duals systems integrable?
\end{itemize}
Whether a system is integrable is a difficult question in itself. To rigorously prove integrability for maps (in the sense of Liouville-Arnold \cite{Ves}, or noncommutative integrability \cite{LGPV,MF}) one needs to provide a symplectic structure together with sufficiently many Casimirs and integrals in involution.

We do not take this route in this paper, instead we employ the notion of complexity, as measured by growth of degrees, as follows: For a given map one can define an integer sequence $\{d_n\}_{n=0}^\infty$ where $d_n$ denotes the degree of the $n$th iterate of the map. Here it suffices to take the degree of the numerator of the last component of the $n$th iterate. Also, to enable the calculation of sufficiently many iterates, we start with initial values which are affine linear functions of one variable, e.g.
\[
u_i=a_{i1}+a_{i2}z,\qquad v_i=a_{i3}+a_{i4}z,\qquad i\in\{1,2\},\qquad a_{ij}\in\mathbb{Z}.
\]
According to the {\em degree growth conjecture} \cite{FV,HV} we have
\begin{itemize}
\item growth is linear in $n$ $\implies$ equation is linearizable.
\item growth is polynomial in $n$ $\implies$ equation is integrable.
\item growth is exponential in $n$ $\implies$ equation is non-integrable.
\end{itemize}
Needless to say that for periodic maps the sequence $d_n$ does not grow.

Methods have been developed to determine an upperbound on the growth of degrees \cite{V}, and to obtain exact formulae \cite{HHVQ, H, RGWM}. In practice, a good indication of the degree growth formula (in the case of polynomial growth) can often be obtained from the first 20 terms of the sequence. Note that in the case of exponential growth it may not practically be possible to iterate the map 20 times.

The remainder of this paper is organised as follows. It begins with the formulation of a system of two $d$-th order ordinary difference equations (O$\Delta$Es) in Section \ref{Formulation}.  Then a procedure to construct systems that preserve the same integral is introduced.  We also provide conditions for the existence and non existence of dual systems.

As a first example, we consider the Nonlinear Schr\"odinger system of P$\Delta$Es. Since all its integrals are multi-linear in the initial values our procedure produces no dual systems. The next example is a system of O$\Delta$Es derived as a periodic reduction from the two-component potential Korteweg-De Vries (pKdV) system of P$\Delta$Es. For a particular linear combination of integrals we obtain a closed chain of four dual systems. The original equation is equal to a composition of its dual systems, which are periodic maps. For other combinations of integrals an infinite chain of dual systems is obtained. A growth of degree argument indicates that the first dual is linearizable, the second dual is integrable, and the third dual is non-integrable.

Two-component O$\Delta$Es can also be obtained by performing a $(s_1,s_2)$-periodic reduction of a scalar lattice equation where the greatest common divisor of $s_1$ and $s_2$ is 2. As an example, we consider the $(2,4)$-reduction of the lattice modified Korteweg-De Vries scalar equation. The duality structure and their integrability are similar to the those for the pKdV system.

Finally, in Section \eqref{BoussinesqSystem} we apply the procedure to a system of O$\Delta$Es which is derived from the modified-Boussinesq system of P$\Delta$Es.  This system is derived using the standard
staircase of $(d+1,-1)$-type for arbitrary $d\in \mathbb{N}$; the above example, system (\ref{intexa}), corresponds to $d=2$. Using a particular integral we obtain (for any $d$) a closed chain of six dual systems. One dual is related to the original map, another is a periodic map, and the remaining three are related to each other and provide a $d$-dimensional generalisation of an alternating QRT map. Their degree growth indicates integrability.

We note that all of the above mentioned P$\Delta$Es are multi-dimensionally consistent \cite{ABS,NW} and their Lax pairs can be found in \cite{Bridgman}. The integrals for their periodic reductions, which may give rise to dual equations, are obtained from the trace of the so-called monodromy matrix \cite{Papageorgiou,QuispelCapelPapageorgiouNijhoff,vanderKamp}, which is the inversely ordered product of the Lax matrices along a staircase.

\section{Formulation of the Problem  \label{Formulation}}
Let $S$ be a field. We consider the orbit of a point in a $2d$-dimensional ($d \in \mathbb{N}$) space $M=S^{2d}$, with coordinates at time $t$: $(\boldsymbol{u}, \boldsymbol{v})_t = (u_t, \ldots, u_{t+d-1}, v_t, \ldots, v_{t+d-1})$.  Let $f_1 : M \longrightarrow S$ and $g_1 : M \longrightarrow S$ be two scalar functions, and define a function $\boldsymbol{F} : M \rightarrow M$,
\[
F(\boldsymbol{u}, \boldsymbol{v}) = \left( u_2, \ldots, u_{d}, f_1(\boldsymbol{u}, \boldsymbol{v}), v_2, \ldots, v_{d}, g_1(\boldsymbol{u}, \boldsymbol{v}) \right).
\]
We define a discrete dynamical system:
\[
(\boldsymbol{u}, \boldsymbol{v})_{t+1} = \boldsymbol{F}\left( (\boldsymbol{u}, \boldsymbol{v})_{t} \right),
\] where $t \in \mathbb{N}$.   Or alternatively, we write the discrete dynamical system as a mapping:
\[
(\boldsymbol{u}, \boldsymbol{v}) \mapsto (\mapping{\boldsymbol{u}}, \mapping{\boldsymbol{v}})
\]
with
\begin{equation} \label{GeneralTwoCoupledSystem}
  \begin{array}{lcl}
   \mapping{u_1} & = & u_2, \ldots, \mapping{u_{d-1}} =  u_d, \quad  \mapping{u_d} = f_1(\boldsymbol{u}, \boldsymbol{v}) \\[.25cm]
   \mapping{v_1} & = & v_2 , \ldots , \mapping{v_{d-1}} =  v_d,  \quad  \mapping{v_d}  =  g_1(\boldsymbol{u}, \boldsymbol{v}),
  \end{array}
  \end{equation}  where the upshifted index is denoted by the prime.

The system \eqref{GeneralTwoCoupledSystem} is a system of two coupled $d$-th order difference equations.   This type of system might be derived as a reduction of a system of partial difference equations, using a standard staircase of $(s_1,s_2)$-type, with co-prime $s_1, s_2 \in \mathbb{Z}$ (for detailed explanation of the staircase method, see \cite{vanderKamp} and reference in there).  The same type of system can also be derived from a single partial difference equation (scalar equation) using a $(s_1,s_2)$ standard staircase with gcd$(s_1, s_2)=2$. \\\\
We assume that \eqref{GeneralTwoCoupledSystem} has $n$ integrals, i.e.:  $H_k: M \longrightarrow S$, $k = 1, 2, \ldots, n$ for some $n \le 2d$.     Consider the linear combination:
  \[
     H = \sum \limits_{k=1}^n \alpha_k H_k,
  \]
   where $\alpha_k\in S$, $k = 1, 2, \ldots, n$, which is also an integral of \eqref{GeneralTwoCoupledSystem}.  Writing $\mapping{\boldsymbol{u}} = (u_2, \ldots, u_d, u_{d+1})$ and $\mapping{\boldsymbol{v}} = (v_2, \ldots, v_d, v_{d+1})$, we define:
  \[
    \mathcal{H}(u_{d+1}, v_{d+1}) = H(\mapping{\boldsymbol{u}} , \mapping{\boldsymbol{v}} ) - H(\boldsymbol{u}, \boldsymbol{v}).
  \]  We remark that the function $\mathcal{H}$ also depends on $\boldsymbol{u}$ and $\boldsymbol{v}$, however we regard these variables as parameters in the function. \\\\
%
%
In this paper we propose a procedure for constructing a family of dual systems of ordinary difference equations to  \eqref{GeneralTwoCoupledSystem} by exploiting the relation $\mathcal{H}(u_{d+1}, v_{d+1})= 0$.  Since $H$ is an integral for \eqref{GeneralTwoCoupledSystem}, then we have:
\[
   \mathcal{H}\left(f_1(\boldsymbol{u}, \boldsymbol{v}),  g_1(\boldsymbol{u}, \boldsymbol{v}) \right) = 0.
\]  Thus, the existence of solutions for  $\mathcal{H}(u_{d+1}, v_{d+1})= 0$ presents no problems. In most examples, and certainly all of the examples treated in this paper, the function $H$ is a rational function.  This implies that $\mathcal{H}$ is also a rational function, and we denote its numerator by $\mathcal{N}$. The degree of $\mathcal{N}$ in the variables $u_{d+1}$ and $v_{d+1}$ is important.

\begin{proposition}
If the numerator $\mathcal{N}$ of $\mathcal{H}$ is a biquadratic function of  $u_{d+1}$ and $v_{d+1}$, i.e.:
\begin{equation} \label{Numerator}
  \begin{array}{c}
   \mathcal{N}(u_{d+1}, v_{d+1})  =   \mathcal{A} {u_{d+1}}^2 {v_{d+1}}^2 +  \left( \mathcal{B}_1{u_{d+1}}^2  v_{d+1} + \mathcal{B}_2 u_{d+1} {v_{d+1}}^2  \right) \\[.25cm]
                +    \left( \mathcal{C}_1 {u_{d+1}}^2   + \mathcal{C}_2 {u_{d+1}} {v_{d+1}} + \mathcal{C}_3 {v_{d+1}}^2 \right)  +
                     \mathcal{D}_1 u_{d+1} + \mathcal{D}_2 v_{d+1}  +  \mathcal{E},
  \end{array}
\end{equation} where $\mathcal{A}$,  $\mathcal{B}_1$, $\mathcal{B}_2$, $\mathcal{C}_1$, $\mathcal{C}_2$, $\mathcal{C}_3$,  $\mathcal{D}_1$, $\mathcal{D}_2$, and $\mathcal{E}$ are polynomials in $\boldsymbol{u}$ and $\boldsymbol{v}$, then the system  \eqref{GeneralTwoCoupledSystem} has no dual system if and only if
\begin{equation} \label{C1}
\mathcal{A}{f_1}^2 +  \mathcal{B}_2 f_1 +  \mathcal{C}_3 =  \mathcal{A}{g_1}^2 +  \mathcal{B}_1 g_1 +  \mathcal{C}_1  = 0,
\end{equation}
or
\begin{equation} \label{C2}
    \mathcal{C}_1 = \frac{ {\mathcal{B}_1}^2}{4\mathcal{A}},\quad
     \mathcal{C}_2 = \frac{\mathcal{B}_1 \mathcal{B}_2}{\mathcal{A}},\quad \mathcal{C}_3 = \frac{{\mathcal{B}_2}^2}{4 \mathcal{A}},\quad    \mathcal{D}_1 = \frac{{\mathcal{B}_1}^2 {\mathcal{B}_2} }{4 \mathcal{A}^2},\quad  \mathcal{D}_2 = \frac{\mathcal{B}_1 {\mathcal{B}_2}^2 }{4 \mathcal{A}^2},\quad
     \mathcal{E} = \frac{{\mathcal{B}_1}^2 {\mathcal{B}_2}^2}{16 \mathcal{A}^3}.
\end{equation}
\end{proposition}

This is shown is appendix B.

\begin{figure}[b]

\begin{tikzpicture}[->,>=stealth',shorten >=1pt,auto,node distance=3cm,
                    semithick]
  \tikzstyle{every state}=[draw=none]

  \node[state] (A)                    {$(f_1, g_1)$};
  \node[state]         (B) [below left of=A] {$(f_1, g_2)$};
  \node[state]         (C) [below right of=A] {$(f_2, g_1)$};
  \node[state]         (D) [below  of=B] {$(f_3, g_2)$};
  \node[state]         (E) [below of=C]       {$(f_2, g_3)$};
   \node[state]         (F) [below  of=D] {$(f_{2k-1}, g_{2k})$};
  \node[state]         (G) [below of=E]       {$(f_{2k}, g_{2k-1})$};
  \node[state]         (H) [below  of=F] {$(f_{2k+1}, g_{2k})$};
  \node[state]         (I) [below of=G]       {$(f_{2k}, g_{2k+1})$};

  \path (A) edge [left ] node {   $ \mathcal{H}(f_1, v_{d+1}) = 0 $ } (B)
  (A) edge [ right]   node {  $\;\; \mathcal{H}(u_{d+1},g_1) = 0 $  } (C)
  (B)  edge [left ] node {  $\mathcal{H}(u_{d+1},g_2) = 0$ } (D)
  (C)  edge [right ] node {  $\mathcal{H}(f_2, v_{d+1}) = 0$ } (E)
  (D)  edge [left, dotted ] (F)
  (E)  edge [right, dotted ] (G)
  (F)  edge [left ] node {  $\mathcal{H}(u_{d+1}, g_{2k}) = 0$ } (H)
  (G)  edge [right ] node {  $\mathcal{H}(f_{2k},v_{d+1}) = 0$ } (I)
  (H) edge [above] node{stop if equal} (I)
  (I) edge (H);
\end{tikzpicture}
\caption{\label{SchematicProcedure} Schematic representation of the procedure for the computation of a family of dual systems.}
\end{figure}
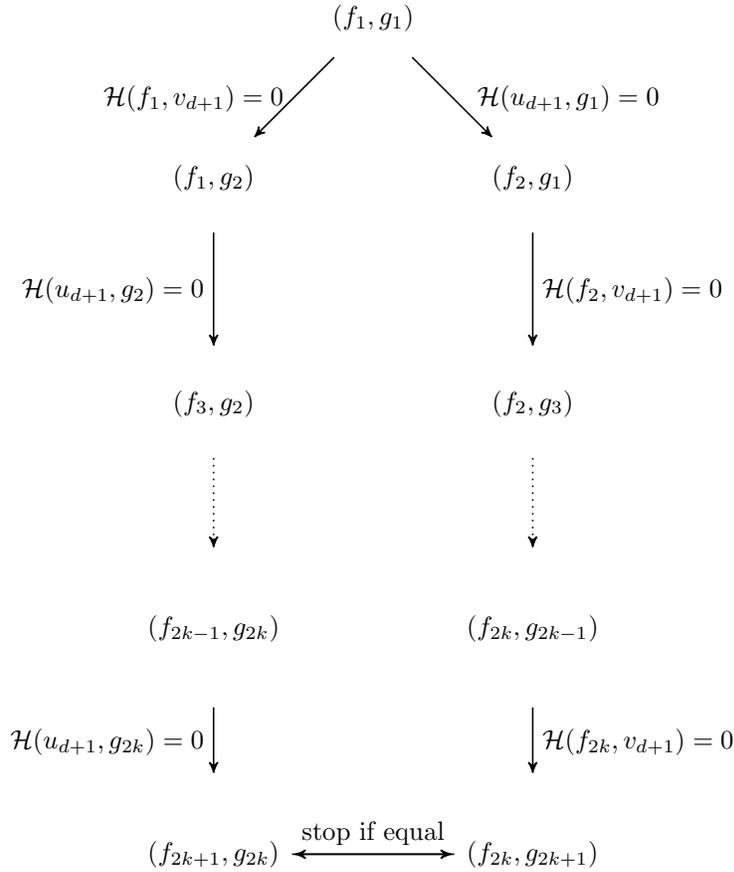

\begin{procedure} \label{proc}
Let the numerator $\mathcal{N}$ of $\mathcal{H}$ be a biquadratic function of the variables $u_{d+1}$ and $v_{d+1}$, such that neither (\ref{C1}) nor (\ref{C2}).
Consider the equation for $v_{d+1}$,
\[
    \mathcal{H}(f_1(\boldsymbol{u},\boldsymbol{v}), v_{d+1}) = 0.
\]   Apart from $v_{d+1} = g_1(\boldsymbol{u}, \boldsymbol{v})$, there exist another solution,  i.e. $v_{d+1} = g_2(\boldsymbol{u}, \boldsymbol{v})$.
Next, we solve
\[
   \mathcal{H}( u_{d+1}, g_1(\boldsymbol{u},\boldsymbol{v})) = 0,
\] for $u_{d+1}$, and we denote the other solution by $u_{d+1} = f_2(\boldsymbol{u}, \boldsymbol{v})$.  Thus, we have constructed two systems of ordinary difference equations:
\begin{equation} \label{Dual1}
  \begin{array}{lcl}
   \mapping{u_1} & = & u_2, \ldots, \mapping{u_{d-1}} =  u_d, \quad  \mapping{u_d} = f_1(\boldsymbol{u}, \boldsymbol{v}) \\[.25cm]
   \mapping{v_1} & = & v_2 , \ldots , \mapping{v_{d-1}} =  v_d,  \quad  \mapping{v_d}  =  g_2(\boldsymbol{u}, \boldsymbol{v}),
  \end{array}
  \end{equation}  and
  \begin{equation} \label{Dual2}
  \begin{array}{lcl}
   \mapping{u_1} & = & u_2, \ldots, \quad  \mapping{u_d} = f_2(\boldsymbol{u}, \boldsymbol{v}) \\[.25cm]
   \mapping{v_1} & = & v_2 , \ldots ,  \quad  \mapping{v_d}  =  g_1(\boldsymbol{u}, \boldsymbol{v}),
  \end{array}
  \end{equation} both having $H$ as their integral.  These systems of equations are both dual to \eqref{GeneralTwoCoupledSystem}.

  We can repeat the process. Let $k \in \{2, 3, \ldots \}$, arbitrary but fixed; suppose that the functions $f_k(\boldsymbol{u}, \boldsymbol{v})$ and $g_k(\boldsymbol{u}, \boldsymbol{v})$ have been computed , and that $\left(\mathcal{A}{f_k}^2 +  \mathcal{B}_2 f_k +  \mathcal{C}_3\right)\left(  \mathcal{A}{g_k}^2 +  \mathcal{B}_1 g_k +  \mathcal{C}_1\right)$ does not vanish. Then, by substituting $u_{d+1} = f_k(\boldsymbol{u}, \boldsymbol{v})$ into $\mathcal{H}(u_{d+1}, v_{d+1})  = 0$, we can solve it for $v_{d+1}$ to construct $v_{d+1} = g_{k+1}(\boldsymbol{u},\boldsymbol{v})$.  Similarly, by substituting $v_{d+1} = g_k(\boldsymbol{u}, \boldsymbol{v})$ into $\mathcal{H}( u_{d+1},  v_{d+1})  = 0$ and solving it for $u_{d+1}$ we get $u_{n+1} = f_{k+1}(\boldsymbol{u}, \boldsymbol{v})$.  It is necessary to check if $f_k = f_{k+1}$ and $g_k = g_{k+1}$, for then the procedure stops.  We illustrate this procedure in Figure \ref{SchematicProcedure}.
  \end{procedure}

If the numerator of the integral is a bilinear function in $u_{d+1}$ and $v_{d+1}$, then the numerator of $\mathcal{H}(u_{d+1}, v_{d+1}) = 0$ can be written as
\[
   \mathcal{A} {u_{d+1}} + \mathcal{B} {v_{d+1} } + \mathcal{C} {u_{d+1} v_{d+1} } + \mathcal{D} = 0,
\] where  $\mathcal{A}$, $\mathcal{B}$, $\mathcal{C}$, and $\mathcal{D}$ are functions of $\boldsymbol{u}$ and $\boldsymbol{v}$.  Then, given $u_{n+1} = f$ (or $v_{n+1} = g$), there is a unique $v_{n+1} = g$ (or $u_{n+1} = f$) that satisfy $\mathcal{H}(u_{n+1}, v_{n+1}) = 0$.  Thus, the dual system cannot be constructed using Procedure \ref{proc}. This result can be extended to $m$-component systems. We give an example of a system which has no duals for this reason.

\begin{example}
Consider the lattice Nonlinear Schr\"{o}dinger (NLS) system,
\[
 \left( x_{m,n+1} - x_{m+1,n} \right) \left( y_{m,n} + \frac{1}{x_{m+1,n+1}} \right) = p-q, \qquad \left( y_{m+1,n} - y_{m,n+1} \right) \left( x_{m+1,n+1} + \frac{1}{y_{m,n}} \right) = p-q,
\]
where $(m,n)$ are coordinates of $\mathbb{Z}^2$.
The $(d-1,1)$-reduction of the NLS system is obtained by introducing
\[
u_{m-(d-1)n}=x_{m,n},\qquad v_{m-(d-1)n}=y_{m,n},
\]
which guarantees the periodicity relation $(x,y)_{m,n}=(x,y)_{m+d-1,n+1}$.
The mapping is defined by the shift in $m$ and gives a $2d$-dimensional map of the form (\ref{GeneralTwoCoupledSystem}), where
\[
\mapping{u_d}=u_1-u_2\frac{p-q}{u_2v_d+1},\qquad
\mapping{v_d}=v_1+v_d\frac{p-q}{u_2v_d+1}.
\]
This map has $d$ integrals, which can be obtained using the staircase method employing the Lax pair given in \cite{Bridgman}. They are multi-linear in all variables.

In particular, for $d=2$, the map is
\[
(u_1,u_2,v_1,v_2) \mapsto (u_2,u_1 - \frac{p-q}{u_2 v_2 + 1} u_2, v_2, v_1 + \frac{p-q}{u_2 v_2 + 1} v_2).
\]
and the integrals are given by
\[
H_1 = u_1 v_2 + u_2 v_1, \text{ and } H_2 = u_1 u_2 v_1 v_2 + p u_1 v_2 + q u_2 v_1 + u_1 v_1 + u_2 v_2.
\]
The function
\[
  \mathcal{H} = \sum \limits_1^2 \alpha_i \left( H_i(\mapping{\boldsymbol{u}}, \mapping{\boldsymbol{v}}) - H_i(\boldsymbol{u},\boldsymbol{v}) \right)
\]
is bi-linear in $u_3$ and $v_3$.  Thus, Procedure \ref{proc} does not produce a dual system.
\end{example}

\section{Application to the two-component potential Korteweg-De Vries equations}
Consider the two-component potential Korteweg-De Vries equations:
\begin{equation}
   \begin{array}{lcl}
       \left( x_{l,m} - x_{l+1, m+1} \right) \left( y_{l+1,m} - y_{l,m+1} \right) & = & p^2 - q^2, \\[.25cm]
       \left( y_{l,m} - y_{l+1, m+1}  \right) \left( x_{l+1,m} - x_{l,m+1} \right) & = & p^2 - q^2.
   \end{array}
\end{equation}  A Lax pair of this system is
\begin{eqnarray*}
   L &=& \begin{pmatrix}
             0            &            0                                  &   x_{l,m}        &  p^2 - k^2 - x_{l,m} y_{l+1,m}  \\
             0           &           0                                    &   1        & -y_{l+1,m}                   \\
             y           &    p^2 - k^2 - x_{l+1,m} y_{l,m}   &   0        &   0   \\
             1           &   -x_{l+1,m}                                       &   0        &    0
                \end{pmatrix},\\
   M &=&\begin{pmatrix}
             0            &            0                                  &   x        &  q^2 - k^2 - x_{l,m} y_{l,m+1}  \\
             0            &           0                                   &   1        & -y_{l,m+1}                   \\
             y           &    q^2 - k^2 - x_{l,m+1} y_{l,m}   &   0        &   0   \\
             1           &   -x_{l,m+1}                                       &   0        &    0
               \end{pmatrix}
\end{eqnarray*}
see \cite{Bridgman} Table 5, pp. 542.  After performing a $(2,1)$-reduction,
\[
x_{m,n}=\xi_{m-2n},\qquad
y_{m,n}=\eta_{m-2n},
\]
we derive a $6$-dimensional mapping, which can be reduced to a $4$-dimensional mapping (by introducing variables $u_1=\xi_2-\xi_1$,
$u_2=\xi_3-\xi_2$, $v_1=\eta_2-\eta_1$, $v_2=\eta_3-\eta_2$),
\begin{equation} \label{2ComponentPKdV}
\left( u_1,u_2,v_1,v_2 \right) \rightarrow \left( u_2, f(\boldsymbol{u},\boldsymbol{v}), v_2,  f(\boldsymbol{v},\boldsymbol{u}) \right),\qquad  f(\boldsymbol{u},\boldsymbol{v}):=-u_1-u_2+ \frac{p^2 - q^2}{v_2}.
 \end{equation}
   By computing the trace of the square of the monodromy matrix we find the following integrals:
\[
H_1  = P - Q,\qquad H_2  = \left( u_1 + u_2 \right) \left(v_1 + v_2 \right) P Q,
\]
  where $P = p^2 - q^2 - u_1 v_2$ and $Q = p^2 - q^2 - u_2 v_1$.

\begin{example} { \bf An example of a closed chain of dual systems}.  \\
Since $I_1$ is bilinear in $u_2$ and $v_2$ then Procedure \ref{proc}, with $H=H_1$, produces no dual system.  However, $H_2$ is biquadratic in $u_2$ and $v_2$ so that we can compute duals with respect to $H_2$ using the Procedure \ref{proc}.  We list the obtained systems in Table \ref{Table2ComponentKdV}.

\begin{table}[h] \renewcommand{\arraystretch}{1.5}
\begin{tabular}{|c|c|c|}  \hline
                  &  Expression for $\mapping{u_2}$ & Expression for $\mapping{v_2}$  \\[.25cm] \hline
 Original Map $\delta_0$ &   $\displaystyle   f(\boldsymbol{u},\boldsymbol{v})  $      &    $\displaystyle f(\boldsymbol{v},\boldsymbol{u})  $ \\[.25cm] \hline
 Dual system $\delta_1$     &  $\displaystyle  f(\boldsymbol{u},\boldsymbol{v}) $      &         $\displaystyle v_1  $  \\[.25cm] \hline
 Dual system $\delta_2$ &  $\displaystyle u_1$  &  $\displaystyle f(\boldsymbol{v},\boldsymbol{u})  $ \\[.25cm] \hline
 Dual system $\delta_3$ & $\displaystyle u_1 $   &   $\displaystyle v_1 $   \\[.25cm] \hline
 \end{tabular}
 \vspace{.5cm}
 \caption{\label{Table2ComponentKdV}  In this table we have listed three duals with respect to $H_2$ to the (2,1)-reduction of the two-component potential Korteweg-De Vries equation.  }
\end{table}

Let us denote the map (\ref{2ComponentPKdV}) by $\delta_0$ and denote the $n$-th dual system by $\delta_n$ ($n=1,\ldots,3$). The map (\ref{2ComponentPKdV}) can be written as a composition of its duals as follows
\[
\delta_0=\delta_1\circ\delta_3\circ\delta_2=\delta_2\circ\delta_3\circ\delta_1.
\]
The involution $\delta_3$ is a reversing symmetry of $\delta_0$, i.e. $\delta_0 \circ \delta_3 \circ \delta_0 = \delta_3$. We also have $\delta_1\iota_1=\iota_1\delta_2$ (and $\delta_i\iota_1=\iota_1\delta_i$ for $i=0,3$), where $\iota_1:(\boldsymbol{u},\boldsymbol{v})\rightarrow (\boldsymbol{v},\boldsymbol{u})$. The duals $\delta_i$, with $i=1,2$, have period 6.
\end{example}

We note that an $n$-dimensional periodic map has $n$ functionally independent integrals. These can be obtained from symmetric functions of the first $n$ iterates
of any function. For the above example, e.g. $\delta_3$ has invariants $u_1+u_2,u_1u_2,v_1+v_2,v_1v_2$ and $\delta_1$ has invariants $u_1+u_2,u_1u_2,\left(v_1 + v_2 \right) P Q$, and $u_1u_2(v_1^2+v_1v_2+v_2^2)-(p^2-q^2)(u_1v_1+u_2v_2)$.

\begin{example} { \bf An example of an infinite family of dual systems}.  \\
Let us now consider (we prefer to denote the coefficients $\alpha,\beta\in S$ instead of $\alpha_1,\alpha_2$)
\[
   H = \alpha H_1 + \beta H_2,
\]
which is an integral for \eqref{2ComponentPKdV}, that is
\[
  \mathcal{H} = H(u_2, u_3, u_4, v_2, v_3, v_4)  - H(u_1, u_2, u_3, v_1, v_2, v_3) = 0,
\] if $u_3=f(\boldsymbol{u},\boldsymbol{v})$ and $v_3=f(\boldsymbol{v},\boldsymbol{u})$.  Following Procedure \ref{proc}, we obtain
\begin{equation} \label{pkdvd1}
\left( u_1,u_2,v_1,v_2 \right) \rightarrow \left( u_2, f(\boldsymbol{u},\boldsymbol{v}), v_2,  v_1 -\frac{\alpha}{\beta(u_1+u_2)P} ) \right)
 \end{equation} as a dual system for \eqref{2ComponentPKdV}, as well as
\begin{equation} \label{pkdvd2}
\left( u_1,u_2,v_1,v_2 \right) \rightarrow \left( u_2, u_1+\frac{\alpha}{\beta(v_1+v_2)Q}, v_2, f(\boldsymbol{v},\boldsymbol{u}) \right).
 \end{equation}
The next dual system, to (\ref{pkdvd1}), is
\begin{equation} \label{pkdvd3}
\left( u_1,u_2,v_1,v_2 \right) \rightarrow \left( u_2, u_1-\frac{\alpha\beta P^2(u_1+u_2)^2}{(\alpha u_2+\beta PQ(u_1+u_2))(\alpha-\beta(u_1+u_2)(v_1+v_2)P)}, v_2,  v_1 -\frac{\alpha}{\beta(u_1+u_2)P} \right),
 \end{equation}
 and a similar expression is obtained for the dual to (\ref{pkdvd2}). After this step, the expression for next dual system becomes too complicated to write down here. The procedure does not seems to end.\\[2mm]

\noindent
Based on 20 iterates of each map we have obtained the following growth formulas. For the original map (\ref{2ComponentPKdV}) we find $d_n=n$.
For the first dual, (\ref{pkdvd1}), we find $d_n\sim 2n$ up to a periodic
sequence with period 3. For the next dual, (\ref{pkdvd3}), we find $d_n\sim \frac{29}{14}n^2-\frac{53}{14}n$ up to a periodic
sequence with period 7. For the next dual (to (\ref{pkdvd3}), not displayed) we could only calculate 4 iterates (in a reasonable amount of time). The sequence $d_n=1, 20, 102, 358, 1189$ seems to grow exponentially fast.
\end{example}

\section{Application to the modified Korteweg-De Vries scalar equation\label{mKdvSystem}}
Consider the modified Korteweg-De Vries scalar equation (also known as $H_3$ in the Adler-Bobenko-Suris classification \cite{ABS}),
\begin{equation}\label{mKdVPDE}
p\left( x_{m,n}x_{m+1,n} + x_{m,n+1} x_{m+1,n+1}\right) - q \left( x_{m,n} x_{m,n+1} + x_{m+1,n} x_{m+1,n+1} \right)+  \delta \left(p^2 - q^2 \right) = 0,
\end{equation} on a two dimensional lattice $\mathbb{Z}^2$. For $\delta = 0$, the Lax matrices for this equation are, see \cite[Table 1, pp. 522]{Bridgman},
\[
  L =  \frac{1}{\sqrt{x_{m,n} x_{m+1,n}}}\left( \begin{array}{cc}
        k x_{m,n} &  - p x_{m,n} x_{m+1,n} \\
        p & - k x_{m+1,n}
   \end{array}\right), \quad M =    \frac{1}{\sqrt{x_{m,n} x_{m,n+1}}}\left( \begin{array}{cc}
        k x_{m,n} &  - q x_{m,n} x_{m,n+1} \\
        q & - k x_{m,n+1}
   \end{array}\right).
\]

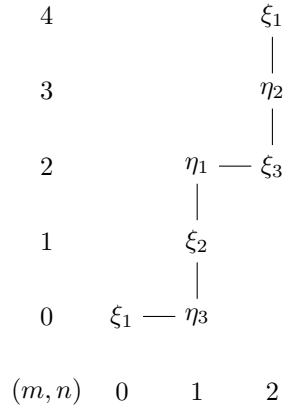
\begin{figure}[b]
\begin{tikzpicture}[node distance=1cm]

  \node(A)                    {$\xi_{1}$};
  \node(B) [below of=A] {$\eta_{2}$};
  \node(C) [below of=B] {$\xi_{3}$};
  \node(D) [left  of=C] {$\eta_{1}$};
  \node(E) [below of=D]       {$\xi_{2}$};
   \node(F) [below  of=E] {$\eta_{3}$};
  \node(G) [left of=F]       {$\xi_{1}$};

  \node(G1) [below of=G] {$0$};
  \node(G2) [right of=G1] {$1$};
  \node(G3) [right of=G2] {$2$};
  \node(G4) [left of=G1] {$(m,n)$};
  \node(G5) [above of=G4] {$0$};
  \node(G6) [above of=G5] {$1$};
  \node(G7) [above of=G6] {$2$};
  \node(G8) [above of=G7] {$3$};
  \node(G9) [above of=G8] {$4$};

  \draw(A) -- (B);
  \draw(B) -- (C);
  \draw(C) -- (D);
  \draw(D) -- (E);
  \draw(E) -- (F);
  \draw(F) -- (G);

\end{tikzpicture}
%
\caption{\label{Staircase} The standard staircase for $(2,4)$-reduction.        }
\end{figure}

According to the $(2,4)$-staircase, see Figure \ref{Staircase}, we define the following reduction:
\[ \begin{array}{lllllllll}
          x_{m,n}    & \longmapsto & \xi_1,   & x_{m+1,n}      & \longmapsto & \eta_3,  & x_{m+1,n+1} & \longmapsto & \xi_2,  \\
     x_{m+1,n+2} &  \longmapsto & \eta_1,  & x_{m+2,n+2}  & \longmapsto & \xi_3,  & x_{m+2,n+3} & \longmapsto & \eta_2.  \\
    \end{array}
\]
Then, in terms of variables $u_1=\xi_1/\xi_2$, $u_2=\xi_2/\xi_3$, $v_1=\eta_1/\eta_2$, $v_2=\eta_2/\eta_3$ the lattice equation \eqref{mKdVPDE} reduces to the mapping
\begin{equation} \label{SystemmKdVODE}
\left( u_1,u_2,v_1,v_2 \right) \rightarrow \left( u_2, f(\boldsymbol{u},\boldsymbol{v}), v_2,  f(\boldsymbol{v},\boldsymbol{u}) \right),\qquad f(\boldsymbol{u},\boldsymbol{v}) = \frac{ p v_2-q}{u_1u_2(p-qv_2)}.
\end{equation}

Computing the trace of the monodromy matrix along the standard $(2,4)$-staircase, we find two functionally independent integrals:
\begin{equation} \label{IntegralmKdV1}
\begin{array}{lcl}
    H_1 (\boldsymbol{u}, \boldsymbol{v})  & = &
 \left( v_{{1}}u_{{1}}-{\frac {u_{{1}}}{v_{{2}}}}+v_{{2}}v_{{1}}+{
\frac {1}{v_{{2}}u_{{2}}}}+{\frac {1}{u_{{2}}u_{{1}}}}-{\frac {v_{{2}}
}{u_{{1}}}}-{\frac {u_{{2}}}{v_{{1}}}}+v_{{2}}u_{{2}}+{\frac {1}{v_{{1
}}u_{{1}}}}+u_{{2}}u_{{1}}-{\frac {v_{{1}}}{u_{{2}}}}+{\frac {1}{v_{{2
}}v_{{1}}}} \right) {p}^{2}\\
&&+ \left( -{\frac {1}{v_{{1}}v_{{2}}u_{{1}}}
}+\frac{1}{u_2}+\frac{1}{v_2}-{\frac {1}{v_{{1}}v_{{2}}u_{{2}}}}-v_{
{1}}u_{{1}}u_{{2}}+\frac{1}{u_1}+v_{{2}}-v_{{1}}v_{{2}}u_{{2}}-v_{{2}
}u_{{1}}u_{{2}}-{\frac {1}{v_{{1}}u_{{1}}u_{{2}}}}\right.\\
&&\left.+v_{{1}}-{\frac {1}{
v_{{2}}u_{{1}}u_{{2}}}}+u_{{1}}-v_{{1}}v_{{2}}u_{{1}}+u_{{2}}+{v_{{1}}
}^{-1} \right) qp+ \left( v_{{1}}v_{{2}}u_{{1}}u_{{2}}+{\frac {1}{v_{{
1}}v_{{2}}u_{{1}}u_{{2}}}} \right) {q}^{2}
  \\
 H_2 (\boldsymbol{u}, \boldsymbol{v})   & = &
    \left( -{\frac {v_{{2}}v_{{1}}}{u_{{2}}u_{{1}}}}-{\frac {u_{{2}}u_{{1
}}}{v_{{2}}v_{{1}}}} \right) {p}^{2} + \left( {\frac {u_{{2}}}{v_{{2}}v
_{{1}}}}+{\frac {v_{{2}}v_{{1}}}{u_{{2}}}}+{\frac {v_{{2}}v_{{1}}}{u_{
{1}}}}+{\frac {u_{{1}}}{v_{{2}}v_{{1}}}}-\frac{1}{u_2}-\frac{1}{v_2}
+{\frac {u_{{2}}u_{{1}}}{v_{{2}}}}-\frac{1}{u_1}\right.\\
&&\left.-v_{{2}}+{\frac {v_{{
2}}}{u_{{2}}u_{{1}}}}+{\frac {v_{{1}}}{u_{{2}}u_{{1}}}}-v_{{1}}-u_{{1}
}+{\frac {u_{{2}}u_{{1}}}{v_{{1}}}}-u_{{2}}-\frac{1}{v_1} \right) qp \\
&&+
 \left( -v_{{2}}v_{{1}}+v_{{1}}u_{{2}}-{\frac {v_{{1}}}{u_{{1}}}}+v_{{
2}}u_{{1}}-{\frac {v_{{2}}}{u_{{2}}}}-u_{{2}}u_{{1}}-{\frac {1}{u_{{2}
}u_{{1}}}}-{\frac {u_{{2}}}{v_{{2}}}}+{\frac {1}{v_{{2}}u_{{1}}}}-{
\frac {u_{{1}}}{v_{{1}}}}+{\frac {1}{v_{{1}}u_{{2}}}}-{\frac {1}{v_{{2
}}v_{{1}}}} \right) {q}^{2}.
 \end{array}
\end{equation}

\begin{example} { \bf Another closed chain of dual systems}.  \\

\noindent
We consider a special linear combination of the integral $H_1$ and $H_2$,
\begin{equation} \label{IntegralRelationmKdV}
   H(\boldsymbol{u},\boldsymbol{v}) = H_1(\boldsymbol{u},\boldsymbol{v})  + H_2(\boldsymbol{u},\boldsymbol{v}).
\end{equation}
Applying the procedure \ref{proc}, we derive three systems which are dual to \eqref{SystemmKdVODE}. They are presented in Figure \ref{TablemKdV}.

\begin{table}[h] \renewcommand{\arraystretch}{2.25}
\begin{tabular}{|c|c|c|}  \hline
                  &  Expression for $\mapping{u_2}$ & Expression for $\mapping{v_2}$  \\[.25cm] \hline
 Original Map $\delta_0$ &   $\displaystyle   f(\boldsymbol{u},\boldsymbol{v})  $      &    $\displaystyle f(\boldsymbol{v},\boldsymbol{u})  $ \\[.25cm] \hline
 Dual system $\delta_1$ &  $\displaystyle  f(\boldsymbol{u},\boldsymbol{v}) $      &         $\displaystyle v_1  $  \\[.25cm] \hline
 Dual system $\delta_2$ &  $\displaystyle u_1$  &  $\displaystyle f(\boldsymbol{v},\boldsymbol{u})  $ \\[.25cm] \hline
 Dual system $\delta_3$ & $\displaystyle u_1 $   &   $\displaystyle v_1 $   \\[.25cm] \hline
 \end{tabular}
 \vspace{.5cm}
 \caption{\label{TablemKdV}  In this table, which is the same table as \ref{Table2ComponentKdV} but with different function $f(\boldsymbol{u},\boldsymbol{v})$, we listed four system which are dual to each other for the $(2,4)$-reduction of mKdV system with $H$ given by \eqref{IntegralRelationmKdV}.  }
\end{table}

Denoting the map (\ref{SystemmKdVODE}) by $\delta_0$ and the $n$-th dual system by $\delta_n$ ($n=1,\ldots,3$), we can again write the original map as a composition of its duals,
\[
\delta_0=\delta_1\circ\delta_3\circ\delta_2=\delta_2\circ\delta_3\circ\delta_1.
\]

As in the previous section the duals are periodic, we have $\delta_1^6=\delta_2^6=\delta_3^2=id$.
\end{example}

\begin{example} { \bf An infinite chain of dual systems}.  \\
Starting from a general linear combination, with $\alpha,\beta\in S$,
\[
  H = \alpha H_1(\boldsymbol{u},\boldsymbol{v})  + \beta H_2(\boldsymbol{u},\boldsymbol{v}),
\] our procedure \ref{proc} gives rise to a hierarchy of infinitely many systems of O$\Delta$Es. We list the first couple:
\begin{equation}\label{dual11mkdv}
\left( u_1,u_2,v_1,v_2 \right) \rightarrow \left( u_2, f(\boldsymbol{u},\boldsymbol{v}), v_2, w(\boldsymbol{v},\boldsymbol{u}) ) \right)	
\end{equation}
and
\begin{equation}\label{dual12mkdv}
\left( u_1,u_2,v_1,v_2 \right) \rightarrow \left( u_2, w(\boldsymbol{u},\boldsymbol{v}) , v_2, f(\boldsymbol{v},\boldsymbol{u}) ) \right)	
\end{equation}
where
\[
	w(\boldsymbol{u},\boldsymbol{v}) =
-{\frac {u_{{1}} \left( \beta\,qu_{{2}}{v_{{1}}}^{2}v_{{2}}-\beta\,pu_
{{2}}v_{{1}}v_{{2}}-\beta\,p{v_{{1}}}^{2}v_{{2}}-\alpha\,qu_{{2}}v_{{1
}}+\alpha\,qv_{{1}}v_{{2}}+\alpha\,pu_{{2}}+\beta\,pv_{{1}}-\alpha\,q
 \right) }{-\alpha\,qu_{{2}}{v_{{1}}}^{2}v_{{2}}+\alpha\,p{v_{{1}}}^{2
}v_{{2}}+\beta\,pu_{{2}}v_{{1}}v_{{2}}+\alpha\,qu_{{2}}v_{{1}}-\alpha
\,qv_{{1}}v_{{2}}-\beta\,pu_{{2}}-\beta\,pv_{{1}}+\beta\,q}}
\]
are the first duals of (\ref{SystemmKdVODE}). The next dual system becomes too complicated to write down and the procedure does not seem to end.

Based on 20 iterates for each map, we have obtained that the original map (\ref{SystemmKdVODE}) has $d_n \sim \frac{19}{30}n^2-\frac{n}{2}$ up to a periodic sequence with period 15.
For the duals (\ref{dual11mkdv}) and (\ref{dual12mkdv}), we found $d_n \sim \frac{7}{6}n^2-\frac{7}{6}n$ up to periodic sequences with period 12.
For the next dual (not displayed) the growth seems exponential: $d_n=1, 13, 61, 265, 1097$.
\end{example}

\section{Application to the modified Boussinesq equation \label{BoussinesqSystem}}
Consider the following system of partial difference equations on a two-dimensional lattice  $\mathbb{Z}^2$ depending on two parameters $p$ and $q$,
\begin{equation}
  \begin{array}{lcl}
      x_{m+1,n+1}\left(p y_{m+1,n} - q y_{m,n+1} \right) - y_{m,n} \left( p x_{m,n+1} - q x_{m+1,n} \right) & = & 0 \\
     x_{m,n} y_{m+1,n+1} \left( p y_{m+1,n} - q y_{m,n+1} \right) - y_{m,n} \left( p x_{m+1,n} y_{m,n+1} - q x_{m,n+1} y_{m+1,n} \right) & = & 0.
   \end{array}
\end{equation}
This system of equations, which is known as the modified Boussinesq System (see \cite{Bridgman}), can be derived by computing the compatibility condition
\[
    M_{m+1,n} L_{m,n}  - L_{m,n+1} M_{m,n} = 0
\]
where
\begin{eqnarray*}
  L_{m,n}  &=& \frac{1}{y_{m,n}}  \left(\begin{array}{ccc}
                           p y_{m+1,n} & 0 &  -k \\[.25cm]
                           -k x_{m+1,n} y_{m,n}  & p y_{m,n}  & 0 \\ [.25cm]
                           0  & \displaystyle - k \frac{y_{m,n} y_{m+1,n}}{x_{m,n}} & \displaystyle p \frac{x_{m+1,n} y_{m,n}}{x_{m,n}}
                    \end{array}\right), \\
    M_{m,n}  &=& \frac{1}{y_{m,n}}  \left(\begin{array}{ccc}
                           q y_{m,n+1} & 0 &  -k \\[.25cm]
                           -k x_{m,n+1} y_{m,n}  & q y_{m,n}  & 0 \\ [.25cm]
                           0  & \displaystyle - k \frac{y_{m,n} y_{m,n+1}}{x_{m,n}} & \displaystyle q \frac{x_{m,n+1} y_{m,n}}{x_{m,n}}
                    \end{array}\right).
\end{eqnarray*}

The standard $(d+1,-1)$-staircase reduction yields the following mapping,
\begin{equation} \label{dminusonereductionsystem}
  \begin{array}{lcl}
      \mapping{\xi_1} & = & \xi_2, \quad \ldots, \mapping{\xi_{d}} = \xi_{d+1}, \\[.25cm]
       \mapping{\xi_{d+1}} & = & \displaystyle \frac{ \eta_1 \left( p \xi_{d+1} - q  \xi_2  \right) }{ p\eta_2 - q \eta_{d+1}}, \\[.5cm]
        \mapping{\eta_1} & = & \eta_2, \quad \ldots, \mapping{\eta_{d}} = \eta_{d+1}, \\[.25cm]
        \mapping{\eta_{d+1}} & = &  \displaystyle \frac{ \eta_1 \left( p \xi_2  \eta_{d+1} -  q \xi_{d+1}  \eta_2   \right) }{ \xi_1 \left( p \eta_2 - q  \eta_{d+1} \right) }.
  \end{array}
\end{equation}  It is straightforward to show that the function
\begin{equation} \label{dminusonereductionintegral}
   H := p  \sum \limits_{1}^d \left( \frac{\xi_k}{\xi_{k+1}} + \frac{\eta_{k+1}}{\eta_k} + \frac{\eta_k \xi_{k+1}}{\xi_k \eta_{k+1}} \right) - q \left( \frac{\xi_1}{\xi_{d+1}} + \frac{\eta_{d+1}}{\eta_1} + \frac{\eta_1 \xi_{d+1}}{\xi_1 \eta_{d+1}}\right),
\end{equation} is an integral for the system of ordinary difference equations \eqref{dminusonereductionsystem}.  \\\\
The system \eqref{dminusonereductionsystem} and its integral \eqref{dminusonereductionintegral}  are invariant with respect to scaling: $\xi_k \longmapsto \lambda \xi_k$, $\eta_k \longmapsto \lambda \eta_k$, $k =1, 2, \ldots, d+1$.   Therefore, we can reduce the dimension of the system by two, by using the transformation:
\[
  u_1 = \frac{\xi_1}{\xi_2}, \ldots,  u_d = \frac{\xi_{d}}{\xi_{d+1}}, \; y_1 = \frac{\eta_2}{\eta_1}, \ldots y_d = \frac{\eta_{d+1}}{\eta_d}.
\]  Furthermore, let us define:
\begin{equation} \label{DefinitionofUV}
   U = \prod \limits_{k=2}^d u_k, \text{ and } V = \prod \limits_{k=2}^d v_k.
 \end{equation}    Then, the system of difference equation \eqref{dminusonereductionsystem} becomes:
\begin{equation} \label{originalmap}
   \begin{array}{lcl}
         \mapping{u_1} & = & u_2,  \; \ldots,  \mapping{u_{d-1}} \; = \; u_d, \\[.5cm]
         \mapping{u_d} & = & \displaystyle \frac{ v_1 \left( p  - q V \right) }{ p - qU } =: f_1(\boldsymbol{u},\boldsymbol{v}) \\[.5cm]
            \mapping{v_1} & = & v_2,  \; \ldots,  \mapping{v_{d-1}} \; = \; v_d, \\[.5cm]
         \mapping{v_d} & = & \displaystyle \frac{1}{u_1 v_1 U V} \;\;  \frac{  p UV - q }{ p- q V } =: g_1(\boldsymbol{u},\boldsymbol{v}),
   \end{array}
\end{equation}  where $\boldsymbol{u} = (u_1, \ldots, u_d)$, $\boldsymbol{v} = (v_1, \ldots, v_d)$.  Consequently, the integral \eqref{dminusonereductionintegral} becomes:
\begin{equation}  \label{Integral}
   H := p \left( \sum\limits_{k=1}^d  u_k + v_k + \frac{1}{u_k v_k} \right) - q \left( u_1U  + v_1V  +  \frac{1}{u_1Uv_1V} \right).
\end{equation}  We define the following:
\begin{equation} \label{MainIntegralRelation}
  \mathcal{H}(u_{d+1}, v_{d+1})  =    H(\mapping{\boldsymbol{u}}, \mapping{\boldsymbol{v}}) - H(\boldsymbol{u}, \boldsymbol{v}),
\end{equation} where $\mapping{\boldsymbol{u}} = (u_2, \ldots, u_{d+1})$, and $\mapping{\boldsymbol{v}} = (v_2, \ldots, v_{d+1})$.  Clearly, since $H$ is integral of \eqref{originalmap}, then:
\[
  \mathcal{H}(f_1(\boldsymbol{u}, \boldsymbol{v}), g_1(\boldsymbol{u}, \boldsymbol{v})) = 0.
\]  Following procedure \ref{proc} to compute dual systems, we construct systems of ordinary difference equations which preserve the integral \eqref{Integral}.  The procedure ends after three steps (when $k = 3$).  In Table \ref{TableBousinesq}  we have listed five duals for \eqref{originalmap}.

\begin{table}[h] \renewcommand{\arraystretch}{2.25}
\begin{tabular}{|c|c|c|}  \hline
                  &  Expression for $u_{d+1}$ & Expression for $v_{d+1}$  \\[.25cm] \hline
 Original Map $\delta_0$ &   $\displaystyle   f_1(\boldsymbol{u},\boldsymbol{v}) $      &    $\displaystyle   g_1(\boldsymbol{u},\boldsymbol{v})$ \\[.25cm] \hline
 Dual system $\delta_1$ &  $\displaystyle  f_1(\boldsymbol{u},\boldsymbol{v})   $      &         $ \displaystyle g_2(\boldsymbol{u},\boldsymbol{v}) := f_1(\boldsymbol{v},\boldsymbol{u}) $  \\[.25cm] \hline
 Dual system $\delta_2$ &  $f_2(\boldsymbol{u}, \boldsymbol{v}) := u_1$  &  $\displaystyle g_1(\boldsymbol{u},\boldsymbol{v})$ \\[.25cm] \hline
 Dual system $\delta_3$ & $\displaystyle f_3(\boldsymbol{u},\boldsymbol{v}):=g_1(\boldsymbol{v},\boldsymbol{u}) $   &   $ \displaystyle g_2(\boldsymbol{u},\boldsymbol{v})  $     \\[.25cm] \hline
 Dual system $\delta_4$ &  $f_2(\boldsymbol{u}, \boldsymbol{v}) $    &   $g_3(\boldsymbol{u}, \boldsymbol{v}) := v_1$   \\[.25cm] \hline
 Dual system $\delta_5$  &  $\displaystyle f_3(\boldsymbol{u}, \boldsymbol{v}) $  &    $g_3(\boldsymbol{u}, \boldsymbol{v})$   \\[0.25cm] \hline
 \end{tabular}
 \vspace{.5cm}
 \caption{\label{TableBousinesq} A chain of six systems which are dual with respect to $H$ given by \eqref{Integral}.  The original map is the $(d+1,-1)$-staircase reduction of the modified Boussinesq system. }
\end{table}

\subsection*{Relations between the duals}
Let us denote the map \eqref{originalmap} by $\delta_0$ and denote the $n$-th dual system by $\delta_n$ ($n=1,\ldots,5$). We further introduce the following involutions, which are symmetries of $H$,
\[
\iota_1:(\boldsymbol{u},\boldsymbol{v})\rightarrow (\boldsymbol{v},\boldsymbol{u}),\qquad
\iota_2:(\boldsymbol{u},\boldsymbol{v})\rightarrow (\boldsymbol{u},\frac{1}{\boldsymbol{u}.\boldsymbol{v}}),
\]
where the . indicates component-wise multiplication. We have the following relations:
\[
\delta_0\iota_1=\iota_1\delta_3,\
\delta_1\iota_1=\iota_1\delta_1,\
\delta_2\iota_1=\iota_1\delta_5,\
\delta_4\iota_1=\iota_1\delta_4,
\]
and
\[
\delta_0\iota_2=\iota_2\delta_3,\
\delta_1\iota_2=\iota_2\delta_5,\
\delta_2\iota_2=\iota_2\delta_2,\
\delta_4\iota_2=\iota_2\delta_4.
\]
This shows that the dual $\delta_3$ is equivalent to the original map $\delta_0$, and that duals $\delta_1$, $\delta_2$, and $\delta_5$ are equivalent to each other. Another symmetry of $H$ and a reversing symmetry for the dual maps $\delta_i$ with $i=1,2,5$ (and $i=4$) is given by
\[
\iota_3:(\boldsymbol{u},\boldsymbol{v})\rightarrow (u_d,\ldots,u_2,u_1,v_d,\ldots,v_2,v_1).
\]
We do not know a reversing symmetry for the maps $\delta_0,\delta_3$.

\subsection*{Generalisation of an alternating QRT-map}


For dual system $\delta_5$, the equation for $\boldsymbol{v}$ is decoupled from the rest, so we can consider the subsystem for $\boldsymbol{v}$ independently. All symmetric polynomials of $v_1,\ldots,v_d$ are integrals. The general solution for the $\boldsymbol{v}$-equation is a periodic sequence.

For $d=2$, the solution can be written as
\[
v_n=\frac{v_1+v_2}{2}-(-1)^n\frac{v_1-v_2}{2}.
\]
The equation for $\boldsymbol{u}$ is an alternating map, and can be written as
\[
u_{n+1}u_{n-1}f_3(u_n,n)=(u+{n+1}+u_{n-1})f_2(x_n,n)-f_1(x_n,n),
\]
where $f(u_n,n)=(A_0(n)X_n)\times (A_1X_n)$, with $X_n=(u_n^2,u_n,1)^T$ and
\[
A_0(n)=p\begin{pmatrix}
0 & 1 & 0 \\
1 & v_1+v_2 & \frac{1-(-1)^n}{2v_1}+ \frac{1+(-1)^n}{2v_2} \\
0 & \frac{1+(-1)^n}{2v_1}+ \frac{1-(-1)^n}{2v_2} & 0
\end{pmatrix}
-q\begin{pmatrix}
1 & 0 & 0 \\
0 & v_1v_2 & 0 \\
0 & 0 & \frac{1}{v_1v_2}
\end{pmatrix},
\quad
A_1=\begin{pmatrix}
0 & 0 & 0 \\
0 & 1 & 0 \\
0 & 0 & 0
\end{pmatrix},
\]
which is the root of a QRT map; see \cite{QuispelA}\footnote{There should be a minus sign in front of the first terms in \cite[equations (20) and (25)]{QuispelA}.}. The integral (\ref{Integral}) is
an alternating integral and can be written in terms of the above matrices as
\[
H=\frac{X_n^TA_0(n)X_{n+1}}{X_n^TA_1X_{n+1}}.
\]
\\\\
Thus, this dual to the $(d+1,-1)$-reduction of the modified Boussinesq system is a higher dimensional generalisation of an alternating (root of a) QRT-map.

\subsection*{Growth of degrees, Integrals}
A study of the degree growth of the maps $\delta_i$ ($i\neq 4$) indicates that they are integrable, with quadratic growth.

For the original map $\delta_0$ (and hence for $\delta_3$) integrals can be constructed using the staircase method. The trace of the monodromy matrix produces the integral $H$ (\ref{dminusonereductionintegral}) for $d = 2 \mod 3$. For other values of $d$, e.g. $d=3$ and $d=5$, it can be obtained from the trace of the square of the monodromy matrix. However, for $d=4$ the staircase method only produces three independent integrals; the function $H$ is a fourth independent integral. The function
\begin{align*}
&p\left( \sum _{i=1}^d \sum _{j=1}^i
\frac{\prod_{k=1}^{d-i}u_{k+j-2}u_{k+d+j-1}}{\prod_{k=1}^{i-1}u_{d-k+j-1}}
+
\frac{u_d\prod_{k=0}^{d-i-1}u_{k}u_{d+k+1}}{
\prod _{k=0}^{i-1-j}u_{d-i+1+k}
\prod _{k=0}^{j-1}u_{2\,d-k-1}}
+\frac{u_{d-1}\prod_{k=0}^{d-i-1}u_{2\,d-1-k}u_{d-2-k}}{
\prod _{k=0}^{i-1-j}u_{d+i-2-k}\prod _{k=0}^{j-1}u_k} \right)\\
&-q \left( \sum _{i=1}^{d} \sum _
{j=0}^{d-i} \sum _{t=0}^{d-j-i} {\frac {\prod _{k=0}^{j-
1}u_{{k+t}}\prod _{k=0}^{j}u_{{d+k+t}}\prod _{k=1}^{i+j-2}u_{{d+k+t}}
}{\prod _{k=1+j}^{i+j-1}u_{{k+t}}}}+{\frac {\prod _{k=0}^{j-1}u_{{2\,d
-1-k-t}}\prod _{k=0}^{j}u_{{d-1-k-t}}\prod _{k=1}^{i+j-2}u_{{d-1-k-t}}
}{\prod _{k=j+1}^{i+j-1}u_{{2\,d-1-k-t}}}} \right.\\
&\ \ \ +\left.
\sum _{i=1}^{d} \sum _{j=0}^{d-i+1} \sum _{t=0}^{i-2}
 {\frac {1}{\prod _{k=0}^{d-i}u_{{k+t}}u_{{d+k+1+t}}\prod _{k=1
}^{j}u_{{k+t}}\prod _{k=j}^{d-i}u_{{d+k+t}}}}
 \right),
\end{align*}
where $u_{d+i}=v_i$, is another integral of $\delta_0$ for all $d$. This integral is a generalisation of integrals calculated by the staircase method for particular values of $d$.

For the ($2d$-dimensional) maps $\delta_i$ with $i=1,2,5$ there are at least $d+2$ independent integrals.
We present them for $\delta_1$. There are the $d$ integrals which are the symmetric functions in $u_1v_1,\ldots,u_dv_d$, and
we have
\[
K=p\left( \sum_{i=1}^d u_i+v_i \right)-q(u_1U+v_1V)
\]
and
\begin{align*}
L&=p\left(\sum_{i=1}^{2d}\prod_{j=1}^{d-2} u_{i+j}\right)-q\left( u_1U \sum _{i=0}^{d-2} \left( \prod _{j=1}^{i}u_{{d+j}}\sum _{l=i}^{d-2}
 \left( \prod _{k=2}^{d-l-1}u_{{k+i}}\prod _{m=2\,d-l+1+i}^{2\,d}u_{{m
}} \right)  \right)+ \right.\\
&\qquad \left. +v_1V
\sum _{i=0}^{d-2} \left( \prod _{j=1}^{i}u_{{d+1-j}}\sum _{l=i}^{d-2}
 \left( \prod _{k=2}^{d-l-1}u_{{2\,d+1-k-i}}\prod _{m=2\,d-l+1+i}^{2\,
d}u_{{2\,d+1-m}} \right)  \right)\right),
\end{align*}
where $u_{d+i}=v_i$. The original integral $H$ can be expressed in terms of $K$ and symmetric functions in $u_1v_1,\ldots,u_dv_d$. As $\delta_1$ is anti-measure preserving, we were able to find integrals for particular values of $d$ using the method of Discrete Darboux Polynomials \cite{DDP1,DDP2}. The integral $L$ was extracted from those by generalisation.

\section{Concluding remarks}
In contrast with the concept of dual equation (\cite{Quispel}) which produces a unique dual equation to a given scalar equation with a given number of integrals, our proposed procedure produces in general a chain of dual systems. 
Although the theory works for systems with any number of components, we have only considered 2-component examples here.

Regarding the first question raised in the introduction, Q1, a dual system with respect to a linear combination of integrals, $H=\sum_i \alpha_i H_i$, will depend on the parameters $\alpha_i$. One expects the complexity of the dual to grow with the number of parameters. Indeed, in Examples 3.1 and 4.1, where there is only one (omitted) parameter, the duals are of finite order, whereas in Examples 3.2 and 4.2, where there are two parameters, the complexity of the dual systems is higher.

Regarding question Q2, the expectation is that the process of dualising will increase complexity. This was observed in Example 3.2, where the growth of the original equation is linear $\sim n$, the growth of the first dual system is slightly higher $\sim 2n$, the growth of the second dual system is quadratic, and the growth of the third dual is exponential.
Similarly, in section Example 4.2, the growth of the original system is quadratic $\sim \frac{19}{30}n^2$, the growth of the first dual system is slightly higher $\sim \frac{7}{6}n^2$, and the growth of the second dual is exponential.

We have also observed, that complexity may actually decrease. This is somewhat surprising, and it lead to the chains of dual systems being closed chains. The procedure has given rise to interesting examples of periodic maps as well as integrable maps.

\subsection*{Acknowledgment}
This work was started during the La Trobe-Indonesia Collaboration Workshop (2017) and was further supported by Riset WCU-ITB Kerjasama Internasional 2017: Duality in Discrete Integrable Systems, by the Australian Research Council and by a La Trobe Asia grant.

\appendix

\section{Explicit expression for the coefficients in equation (\ref{cs})}
The coefficients $c_i$ in equation (\ref{cs}) are
\begin{align*}
c_1&=-x_{{0}}y_{{0}}y_{{1}}y_{{2}} \left( px_{{1}}y_{{2}}-qx_{{2}}y_{
{1}} \right)
\left( p x_{{2}} (x_{{0}}y_{{0}}-x_{{1}}y_{{1}})-q x_{{1}} (x_{{0}}y_{{0}}+x_{{2}}y_{{2}}) \right) \\
c_2&=-{y_{{0
}}}^{2} \left( px_{{2}}-qx_{{1}} \right)
\left( p x_{{2}}y_{{2}} ( {x_{{0}}}^{2}y_{{1}}-{x_{{1}}}^{2}y_{{0}})-q x_{{1
}}y_{{1}}({x_{{0}}}^{2}y_{{2}}+{x_{{2}}}^{2}y_{{0}}) \right) \\
c_3&= x_{{2}}y_{{2}} \left( {x_{{0}}}^{2}y_{{0}}y_{{1}}+x_{{0}}x_{{1}}{y_{{1
}}}^{2}-{x_{{1}}}^{2}{y_{{0}}}^{2} \right) p\\
&\ \ \ -x_{{1}}y_{{1}} \left( {x_
{{0}}}^{2}y_{{0}}y_{{2}}+x_{{0}}x_{{2}}{y_{{2}}}^{2}-{x_{{2}}}^{2}{y_{
{0}}}^{2} \right) q \\
c_4&={x_{{0}}}^{2}y_{{0}}y_{{1}}y_{{2}} \left( px_{{1}}y_{{2}}-qx_{{2}}y_{{
1}} \right)\\
c_5&=x_{{1}}x_{{2}}{y_{{0}}}^{2} \left( px_{{2}}-qx_{{1}} \right)\\
c_6&=-x_{{1}}x_{{2}}y_{{0}}
\end{align*}

\section{Conditions for a biquadratic $\mathcal{H}$ to not give rise to a dual equation}
Since $\mathcal{H}(f_1(\boldsymbol{u}, \boldsymbol{v}), g_1(\boldsymbol{u}, \boldsymbol{v})) = 0$, then $\mathcal{N}(f_1(\boldsymbol{u}, \boldsymbol{v}), g_1(\boldsymbol{u}, \boldsymbol{v})) = 0$. Thus:
\[
\begin{array}{r}
   \mathcal{N}(u_{d+1}, v_{d+1})  =   \mathcal{A}\left( {u_{d+1}}^2 {v_{d+1}}^2 - {f_1}^2 {g_1}^2\right) +  \mathcal{B}_1\left( {u_{d+1}}^2  v_{d+1} - {f_1}^2g_1\right) +   \\[.25cm]
                 \mathcal{B}_2 \left(u_{d+1} {v_{d+1}}^2 - f_1 {g_1}^2\right)    +  \mathcal{C}_1 \left( {u_{d+1}}^2 - {f_1}^2\right)   + \mathcal{C}_2 \left( {u_{d+1}} {v_{d+1}} - f_1 g_1\right) + \\[.25cm]
                 \mathcal{C}_3 \left( {v_{d+1}}^2 - {g_1}^2\right)   +   \mathcal{D}_1 \left( u_{d+1} - f_1 \right) + \mathcal{D}_2 \left( v_{d+1} - g_1\right),
  \end{array}
\] where we have used $f_1$ and $g_1$ as short notations for $f_1(\boldsymbol{u}, \boldsymbol{v})$ and $g_1(\boldsymbol{u}, \boldsymbol{v})$, respectively.  By setting $u_{d+1} = f_1$, then: $\mathcal{N}(f_{1}, v_{d+1})  = 0$ implies:
\[
     \left( \left\{ \mathcal{A}{f_1}^2 +  \mathcal{B}_2 f_1 +  \mathcal{C}_3 \right\} \left({v_{d+1}} + {g_1} \right) +
                \left\{  \mathcal{B}_1{f_1}^2  +   \mathcal{C}_2 f_1  + \mathcal{D}_2 \right\} \right)  \left(  v_{d+1} - g_1\right) =0.
\]  Then, if $\mathcal{A}{f_1}^2 +  \mathcal{B}_2 f_1 +  \mathcal{C}_3 = 0$ a new solution for $v_{d+1}$ fail to exists.

Similarly, setting $v_{d+1} = g_1$ we have:
\[
     \left( \left\{ \mathcal{A}{g_1}^2 +  \mathcal{B}_1 g_1 +  \mathcal{C}_1 \right\} \left({u_{d+1}} + {f_1} \right) +
                \left\{  \mathcal{B}_2{g_1}^2  +   \mathcal{C}_2 g_1  + \mathcal{D}_1 \right\} \right)  \left(  v_{d+1} - g_1\right) =0.
\]  Thus, if  $\mathcal{A}{g_1}^2 +  \mathcal{B}_1 g_1 +  \mathcal{C}_1  = 0$ a new solution for $u_{d+1}$ fail to exists.  \\\\
Let us now look at the situation where both $\mathcal{H}( f_1(\boldsymbol{u}, \boldsymbol{v}), v_{d+1}) = 0$ and $\mathcal{H}(u_{d+1}, g_1(\boldsymbol{u}, \boldsymbol{v})) = 0$ have a unique solution: $v_{d+1} = g_1(\boldsymbol{u}, \boldsymbol{v})$ and $u_{d+1}  = f_1(\boldsymbol{u}, \boldsymbol{v})$, respectively.   Then this procedure produces no other system apart from \eqref{GeneralTwoCoupledSystem}.  This is the case for example when the numerator \eqref{Numerator} can be reduced to:
\begin{equation} \label{UniqueSolution}
 \mathcal{N} = p(\boldsymbol{u}, \boldsymbol{v}) \left( u_{d+1} - f_1(\boldsymbol{u}, \boldsymbol{v})\right)^2 \left( v_{d+1} - g_1(\boldsymbol{u}, \boldsymbol{v})\right)^2.
\end{equation}  Consider the numerator \eqref{Numerator}, which can be written as
\[
  \begin{array}{c}
   \mathcal{N}(u_{d+1}, v_{d+1})  =  \displaystyle \mathcal{A} {u_{d+1}}^2 \left( {v_{d+1}}^2 + \frac{\mathcal{B}_1}{\mathcal{A}}  v_{d+1} + \frac{\mathcal{C}_1}{\mathcal{A} } \right)   \\[.25cm]
  \displaystyle   +  \mathcal{B}_2 u_{d+1} \left(  {v_{d+1}}^2  + \frac{\mathcal{C}_2}{ \mathcal{B}_2} {v_{d+1}} + \frac{\mathcal{D}_1}{\mathcal{B}_2}  \right)
                    + \mathcal{C}_3  \left(     {v_{d+1}}^2   +   \frac{\mathcal{D}_2}{\mathcal{C}_3} v_{d+1}  +  \frac{\mathcal{E}}{\mathcal{C}_3} \right).
  \end{array}
\]     By setting:
\[
  \mathcal{C}_1 = \frac{\mathcal{A}\mathcal{D}_1}{\mathcal{B}_2}, \mathcal{C}_2 = \frac{\mathcal{B}_1\mathcal{B}_2}{\mathcal{A}},  \mathcal{C}_3 = \frac{\mathcal{A}\mathcal{D}_2}{\mathcal{B}_1},  \text{ and }   \mathcal{E} = \frac{\mathcal{A}\mathcal{D}_1 \mathcal{D}_2}{\mathcal{B}_1\mathcal{B}_2},
\] we have:
\[
    \mathcal{N}(u_{d+1}, v_{d+1})  = \mathcal{A} \left(  {u_{d+1}}^2  + \frac{\mathcal{B}_2 }{\mathcal{A}} u_{d+1} +  \frac{ \mathcal{D}_2}{\mathcal{B}_1} \right)  \left( {v_{d+1}}^2 + \frac{\mathcal{B}_1}{\mathcal{A}}  v_{d+1} + \frac{\mathcal{D}_1}{\mathcal{B}_2 } \right).
\]  Lastly, by setting:
\[
   \mathcal{D}_1 = \frac{{\mathcal{B}_1}^2 {\mathcal{B}_2} }{4 \mathcal{A}^2}  \text{ and }  \mathcal{D}_2 = \frac{\mathcal{B}_1 {\mathcal{B}_2}^2 }{4 \mathcal{A}^2},
\] then $\mathcal{N}$ can be written in the desired form: \eqref{UniqueSolution}.

\end{document}